\documentclass[3p]{elsarticle}
\usepackage{lineno, hyperref}
\modulolinenumbers[5]
\usepackage{amssymb}
\usepackage{amsmath}
\usepackage{graphicx}
\usepackage{multirow,array}
\usepackage{bm}
\usepackage{color}
\newtheorem{thm}{Theorem}
\newtheorem{lem}[thm]{Lemma}
\newdefinition{rmk}{Remark}
\newproof{pf}{Proof}
\newproof{pot}{Proof of Theorem \ref{thm2}}
\newtheorem{cor}{Corollary}

\begin{document}

\begin{frontmatter}

\title{A note on general epidemic region for infinite regular graphs}
\author[UY]{Unjong Yu}

\author[JC]{Jeong-Ok Choi\corref{mycorrespondingauthor}}
\cortext[mycorrespondingauthor]{Corresponding author}
\ead{jchoi351@gist.ac.kr}

\address[UY]{Department of Physics and Photon Science, Gwangju Institute of Science and Technology, \\ Gwangju 61005, South Korea}
\address[JC]{Division of Liberal Arts and Sciences, Gwangju Institute of Science and Technology, \\ Gwangju 61005, South Korea}

\begin{abstract}
We study the contagion game with the bilingual option on infinite regular graphs introduced and modeled mathematically in [N. Immorlica et al. (2007)]. 
In the reference, Immorlica et al. studied conditions for an innovation to become epidemic over infinite regular trees, the grid, and the infinite thick-lines in terms of payoff enhancement and cost of the bilingual option.
We improved their results by showing that the class of infinite regular trees make an innovation least advantageous to become epidemic considering the whole class of infinite regular graphs. Moreover, we show that any infinite $\Delta$-regular graph containing the infinite $\Delta$-tree structure is also least advantageous to be epidemic. Also, we construct an infinite family of infinite $\Delta$-regular graphs (including the thick $\Delta$-line) that is the most advantageous to be epidemic as known so far.  
\end{abstract}


\end{frontmatter}

\section{Introduction}
``Diffusion of innovations'' studies the diffusion processes of new ideas, technologies, products, or services through a society over time \cite{Rogers83}.
It has attracted a lot of interest in various fields such as anthropology, sociology, political science, and economics \cite{Greenhalgh04,Kiesling12}.
It is not trivial to predict that an innovation will diffuse to a society,
because people decide to adopt the innovation after careful consideration of the benefit from the adoption.
Many studies have reported that people adopt the innovation when the exposure (the ratio of adopters of the innovation in the neighborhood) is larger than a certain level of value - called a threshold \cite{Rogers83,Katz85,Valent96,Centola10}.

A threshold is an exposure value assigned for each agent to take innovation if it did not take innovation yet. 
In the simplest case where every agent has the same number of neighbors ($\Delta$) and a constant threshold ($t$) smaller than $1/\Delta$, the final state is trivial: the whole population will adopt the innovation eventually. In other words, the innovation is spread over the population. If an innovation is spread over the population, then we say that it becomes {\it epidemic}.
On the other hand, if a given threshold is high enough, then the innovation stops diffusing at some time. Morris\cite{Morris} showed that with a threshold bigger than $1/2$ an innovation is never spread over any graph structure and both strategies will coexist eventually. 

Inspired by Morris, Immorlica et al.\cite{Immorlica07} developed a model introducing a bilingual option at a cost $r$, which is compatible with both strategies. In this bilingual model a possible threshold for an innovation to be epidemic over an infinite regular graph depends on $r$, say $t_r$. They studied this problem and determined the maximum value for $t_r$ completely for every $r > 0$ on infinite regular trees, the grid, and the infinite thick-lines. We call this maximum value the {\it contagion threshold}. (See also \cite{Wortman08, Easley10}. For similar problems under different conditions, see \cite{Goyal97, Oyama15}.)

In this paper, we revisit this problem and obtain more results by finding a structure attaining the (universal) minimum of contagion thresholds among the whole class of infinite regular graphs. In section 2, we present basic preliminaries for the problem and our main results. In the section, we show that the contagion threshold of any infinite $\Delta$-regular graph containing the infinite $\Delta$-regular tree structure becomes the universal lower bound for contagion thresholds of all infinite $\Delta$-regular graphs. Also, we construct an infinite family of infinite $\Delta$-regular graphs (including the thick $\Delta$-line) that are the most advantageous to be epidemic as known so far. In section 3, we discuss further directions.

\section{Main Results}
\subsection{Preliminaries}
We consider connected infinite regular graphs unless mentioned otherwise and follow the model and setting introduced in \cite{Immorlica07}. We denote by $V(G)$ the vertex set of $G$.
\smallskip

Our main question is: what kind of graph structure on a population is more advantageous for a new strategy to be spread over the old strategy in the population?

An {\it $A$-$B$ coordination game} on an infinite regular graph $G$ is a pair $(G, q)$, where every vertex gets the payoffs from the payoff matrix in Table 1 while it plays a game with its neighbors.

\smallskip
\begin{table}[ht]
\centering 
\begin{tabular}{c||c|c}
 & $A$ & $B$  \\ \hline
 $A$ & $1-q, 1-q$ & 0, 0    \\
 $B$ & 0, 0 & $q, q$   \\ 
\end{tabular}
\caption{The payoffs in an $A$-$B$ coordination game $(G, q)$ \label{Table1}} 
\end{table}

\smallskip
An {\it initial strategy profile} is a function $f_0: V(G) \longrightarrow \{A, B\}$. We call $S_0 = f_0^{-1}(A)$ an {\it initial set} of a game. 
Let $\alpha = \{v_i\}_{i = 1}^{\infty}$ be a sequence of $V(G)$, where every vertex in $V(G)$ appears at least once in $\alpha$. An assignment $f_{\alpha}:\alpha \longrightarrow \{A, B\}$ in $(G, q)$ is called a {\it profile along $\alpha$} if $f_{\alpha}$ is determined as follows: (1) start with an initial strategy profile $f_0$ and let $f_{\alpha} = f_0$ and (2) for each $i$ in order, update $f_{\alpha}(v_i)$ as the strategy that gives the higher payoff to $v_i$ based on the payoff matrix in Table~\ref{Table1} after playing with its neighbors using the current assignment.

In an $A$-$B$ coordination game $(G, q)$, strategy $A$ becomes {\it epidemic} if there is a sequence $\alpha$ of $V(G)$ and a finite initial set $S_0$ such that the rule for $f_{\alpha}$ forces that for every $v \in V(G)$ there is an index $i$ such that $f_{\alpha}(v_i) = A$. We call $Q$ the {\it contagion threshold} of $G$ if (1) $A$ becomes epidemic in $(G, q)$ for every $q < Q$ and (2) $A$ never becomes epidemic in $(G, q)$ for any $q > Q$.  

\smallskip
It is easy to see that $Q \ge 1/\Delta$ for every infinite regular graph. 
Morris \cite{Morris} proved that the contagion threshold $Q$ is always at most $1/2$ and the sharpness is achieved by {\it thick lines}. 
For even $\Delta$, the {\it thick line} $L_{\Delta}$ is a graph with a vertex set $\mathbf{Z} \times \{ 1, 2, \cdots, \Delta/2 \}$ and an edge set $\{((k, i), (l, j)) \colon |k-l| = 1 \}$. 

\medskip

Our main concern in this paper is the case that there are two incompatible strategies $A$ and $B$, and another option $AB$ that is compatible with $A$ and $B$. (We call $AB$ a bilingual option.) 
Immorlica et al. set this bilingual model using a two-player game called a contagion game \cite{Immorlica07}. 
To prevent a game from being trivial such that the bilingual option $AB$ becomes epidemic in every graph, there needs a cost to pay for choosing $AB$.  

\medskip

To begin with, all the vertices have (old) strategy $B$. Now a new strategy $A$ is introduced to finitely many vertices in $V(G)$, and the game starts along a given sequence. If a vertex takes a particular strategy, then the payoff that the player earns is the total of the payoff the player earns from its neighbors. While playing the game, each vertex at its turn chooses a strategy giving the largest payoff among $A$, $B$, or $AB$.

Given a sequence of vertices in $V(G)$ and for $0 < q < 1$ and $r > 0$, a {\it contagion game} $(G, q, r)$ is a coordination game where every vertex plays a game with its neighbors using the payoff matrix presented in Table~\ref{Table2}. In this paper, as mentioned above we consider a contagion game, where there are three strategies for each vertex to choose: $A, B$, or $AB$. If a vertex chooses one strategy, then the total payoff from this choice is the sum of the payoffs from playing with the neighbors of the vertex. For example, if a vertex chooses $A$, then the payoff it gets by playing with a neighbor having strategy $A$, $B$, and $AB$ (resp.) is $1-q$, $0$, and $\max(q, 1-q) - r$ (resp.).

\medskip

\begin{table}[ht]
\centering 
\begin{tabular}{c||c|c|c}
 & $A$ & $B$ & $AB$ \\ \hline
 $A$ & $1-q, 1-q$ & 0, 0  & $1-q, 1-q-r$   \\
 $B$ & 0, 0 & $q, q$  & $q, q-r$   \\
 $AB$ & $1-q-r, 1-q$ & $q-r, q$ & $\max(q, 1-q)-r, \max(q, 1-q)-r$   
\end{tabular}
\caption{The payoffs in a contagion game $(G, q, r)$ \label{Table2}} 
\end{table}

\medskip

Note that the remaining part after deleting the row and column of $AB$ in Table~\ref{Table2} is the same as the payoff matrix in Table~\ref{Table1}. We mimic the definition for the strategy $A$ to be epidemic. 

\smallskip
An {\it initial strategy profile} is a function $f_0: V(G) \longrightarrow \{A, B\}$. We call $S_0 = f_0^{-1}(A)$ an {\it initial set} of a game. 
Let $\alpha = \{v_i\}_{i = 1}^{\infty}$ be a sequence of $V(G)$, where every vertex in $V(G)$ appears at least once in $\alpha$. An assignment $f_{\alpha}:\alpha \longrightarrow \{A, B, AB\}$ in a contagion game $(G, q, r)$ is called a {\it profile along $\alpha$} if $f_{\alpha}$ is determined as follows: (1) start with an initial strategy profile $f_0$ and let $f_{\alpha} = f_0$ and (2) for each $i$ in order, update $f_{\alpha}(v_i)$ as the strategy that gives higher payoff to $v_i$ based on the payoff matrix in Table~\ref{Table2} after playing with its neighbors using the current assignment.

In a contagion game $(G, q, r)$, strategy $A$ becomes {\it epidemic} if there is a sequence $\alpha$ of $V(G)$ and a finite initial set $S_0$ such that the rule for $f_{\alpha}$ forces that for every $v \in V(G)$ there is an index $i$ such that $f_{\alpha}(v_i) = A$. 

In other words, the strategy $A$ becomes {\it epidemic} in $(G, q, r)$ if there are a finite set $S$ and a sequence of vertices in $V(G) - S$, say $v_1, v_2, v_3, \cdots $, satisfying that for every $v$ in $V(G) - S$ there is an index $k$ such that (1) $v = v_k$ and (2) at the $k$-th turn along the sequence the best strategy for $v$ is $A$ when having started with $A$s for the vertices of $S$ while other vertices had initially $B$. \smallskip

Note that the definition of epidemic status depends on $q$ and $r$ as well as $G$ because it is possible that $A$ is not epidemic for different $q$ and $r$ even with the same graph ($G$), the same finite set ($S$), and the same sequence ($\alpha$). \smallskip

For a fixed $r>0$, we call $Q_r$ the {\it contagion threshold} of $G$ if (1) $A$ becomes epidemic in $(G, q, r)$ for every $q < Q_r$ and (2) $A$ never becomes epidemic in $(G, q, r)$ for any $q > Q_r$. For an infinite regular graph $G$, the {\it epidemic region} denoted $\Omega_G$ is $\{ (q, r) \colon A \textrm{ becomes epidemic in } (G, q, r) \}$. Therefore, the boundary curve of an epidemic region is the points consisting of $(Q_r, r)$ for every $r > 0$. \smallskip

Let $\Omega_{\Delta} = \bigcup_G\Omega_G$, where the union is taken over all infinite $\Delta$-regular graphs. In this paper we focus on determining $\Omega_{\Delta}$ for every $\Delta \ge 2$. 
\medskip

It is known that no vertex can change either from $A$ to $B$, from $A$ to $AB$, or from $AB$ to $B$ when taking its best response in a contagion game $(G, q, r)$ \cite{Immorlica07}. Hence there are only two possible types for $(q, r)$ to be in the epidemic region: (1) $A$ is always the best response for every turn of vertices and (2) $AB$ is the best response for some vertex but finitely many turns later the best strategy for the vertex is eventually $A$. \\

It is useful to apply the concept of {\it blocking structure} and results on blocking structures introduced in \cite{Immorlica07}. We say $(X, Y)$ a {\it non-trivial pair} of disjoint sets if either $X \neq \emptyset$ or $Y \neq \emptyset$. For a contagion game $(G, q, r)$, a non-trivial pair $(S_{AB}, S_B)$ of disjoint subsets of $V(G)$ is called a {\it blocking structure} for $(G, q, r)$ if the pair satisfies the following properties:
\begin{enumerate}
\item for every $v \in S_{AB}$, $\deg_{S_B}(v) > \frac{r}{q}\Delta$,
\item for every $v \in S_B$
\begin{enumerate}
\item $(1-q)\deg_{S_B}(v)+\min(q, 1-q)\deg_{S_{AB}}(v) > (1-q-r)\Delta$ and
\item $\deg_{S_B}(v) + q \deg_{S_{AB}}(v) > (1-q)\Delta,$
\end{enumerate}
\end{enumerate}
where $\deg_{S_{AB}}(v)$ and $\deg_{S_{B}}(v)$ are the number of neighbors of $v$ in $S_{AB}$ and $S_{B}$, respectively. \smallskip

\begin{thm}\cite{Immorlica07}\label{blocking}
For every contagion game $(G, q, r)$, strategy $A$ cannot be epidemic in $(G, q, r)$ if and only if every co-finite set of vertices of $G$ contains a blocking structure.
\end{thm}
\smallskip
Also, in the same reference, the authors showed that $(q, r)$ with $q > 1/2$ cannot be in the epidemic region for any infinite regular graph. Therefore, $\max(q, 1-q) = 1-q$ in Table~\ref{Table1} for payoffs to be epidemic. \medskip

\subsection{Our Results}

We find a sufficient condition for each vertex (i.e. agent) to adopt $A$ as its best strategy in the given game $(G, q, r)$. The condition will be formulated in terms of $r$ and $q$. \smallskip

\begin{lem} Let $G$ be an infinite $\Delta$-regular graph.
Suppose that there is an order of finite subsets of vertices $V_0, V_1, \cdots$ such that
\begin{enumerate}
\item
\[ \textrm{The union of $V_i$'s is $V(G)$, that is, } \ \ \bigcup_{i = 0}^{\infty} V_i = V(G) \textrm{ and }\]
\item Given $\epsilon$ with $0 < \epsilon < 1$, for each $i$ and $v \in V_i$,
\[  d_{V_0 \cup V_1 \cup \cdots \cup V_{i-1}}(v) \ge \epsilon \Delta .\]
\end{enumerate}
Then $(q, r)$ satisfying $r > (1-\epsilon)q$ and $q < \epsilon$ is in the epidemic region for $G$.
\end{lem} \vskip 0.1in

\begin{pf}
We prove that for each $i$ every $v \in V_i$ has $A$ as the best strategy under the given condition. We use induction on $i$. Starting from an initial situation that every vertex uses strategy $B$, we locate strategy $A$ at every vertex in $V_0$. This automatically satisfies the basis. Now assume that the vertices in $V_0\cup V_1 \cdots \cup V_{i-1}$ adopted $A$ as their strategies. For each vertex $u \in V_i$ the payoff of strategy $A$ is $p_1$ which is at least $(1-q)\epsilon\Delta$. The payoff of strategy $B$ is $p_2$ which is at most $(1-\epsilon)\Delta q$. The payoff of strategy $AB$ is $p_1+p_2 - r\Delta$. Hence the vertex $u$ has $A$ as the best strategy on $(G, q, r)$ game whenever $q$ and $r$ satisfy $r\Delta > p_2$ and $p_1 > p_2$. In particular, if $r > (1-\epsilon)q$ and $q < \epsilon$, then $p_1 - p_2 \ge (1-q)\epsilon \Delta - (1-\epsilon)\Delta q = \Delta(\epsilon - q) > 0$. Also, $p_1 - (p_1+p_2-r\Delta) = r\Delta - p_2 > (1-\epsilon)q\Delta - p_2 > (1-\epsilon)q\Delta - (1-\epsilon)\Delta q = 0$.  \vskip 0.1in

After locating strategy $A$ at each vertex in $V_0$ we let vertices in sets $V_1, V_2, \cdots$ play their best strategy. Eventually $A$ becomes epidemic along the order. \qed
\end{pf}
\vskip 0.15in

An infinite regular-tree is a connected, acyclic, and infinite graph each of whose degree is the same. 
As an easy consequence, if we pick a vertex in the infinite $\Delta$-regular tree, then by considering $V_i$ be the set of the vertices at depth $i$ and by letting $\epsilon \le \frac{1}{\Delta}$ the set $\{ (q, r) \colon r \ge \frac{\Delta-1}{\Delta}q, q \le \frac{1}{\Delta}\}$ is contained in $\Omega_{T_{\Delta}}$. In fact, it is not hard to see that $\Omega_{T_{\Delta}} = \{ (q, r) \colon r \ge \frac{\Delta-1}{\Delta}q, q \le \frac{1}{\Delta} \} \cup \{(q, r) \colon 2q + \Delta r \le 1 \}$. 
\vskip 0.25in

\begin{thm}\label{tree}
Let $\Omega_{T_{\Delta}}$ be the epidemic region for the infinite $\Delta$-regular tree $T_{\Delta}$. For every infinite $\Delta$-regular graph $G$, $\Omega_{T_{\Delta}} \subseteq \Omega_G$. In other words, $\Omega_{T_{\Delta}}$ is the minimum epidemic region that any infinite $\Delta$-regular graph possibly has. 
\end{thm} \vskip 0.1in

\begin{pf}
By Theorem \ref{blocking} for any $(q, r)$ in the complement of the epidemic region $\Omega_G$, every co-finite set of $V(G)$ contains a blocking structure $(S_B, S_{AB})$ for $(G, q, r)$. Therefore, $\{(q, r) \colon q, r > 0 \} - \Omega_G$ is the union of $\Omega_1$ and $\Omega_2$, where $\Omega_1$ consists of $(q, r)$ allowing a blocking structure with $S_{AB} \neq \emptyset$ and $\Omega_2$ consists of $(q, r)$ allowing a blocking structure with $S_{AB} = \emptyset$. \vskip 0.15in

Let $(q, r)$ be in$\{(q, r) \colon q, r > 0 \} - \Omega_G = \Omega_1 \cup \Omega_2$, and we consider a blocking structure $(S_B, S_{AB})$ in a co-finite set of $V(G)$. 
\vskip 0.15in

Case 1) $S_{AB} \neq \emptyset$: \\
We let $a_v = \deg_{S_B}(v)$ and $b_v = \deg_{S_{AB}}(v)$ for any $v \in S_B$. Hence, $a_v, b_v \ge 0$ and $a_v + b_v \le \Delta$. 
Now $r$ and $q$ must satisfy that $a_v + qb_v > (1-q)\Delta $ and $(1-q)a_v + qb_v > (1-q-r)\Delta$. If we let $d = \min_{u \in S_{AB}} \deg_{S_B}(u)$, then $d > \frac{r}{q}\Delta$ by the first condition in the definition of a blocking structure. In other words, $r < \frac{d}{\Delta}q$. Since $a_v + b_v \le \Delta$, we let $\Delta = a_v + b_v + t_v$ with $t_v \ge 0$. 
From those inequalities we obtain that $\left(2-\frac{t}{\Delta-a_v}\right)q+\frac{\Delta}{\Delta-a_v}r > 1$ and $q > \frac{\Delta - a_v}{2\Delta-a_v-t_v}$. 
Note that there is $v_0 \in S_B$ with $a=\deg_{S_B}(v_0) < \Delta$. Therefore $\frac{\Delta-a_v}{\Delta} \ge \frac{1}{\Delta}$. The line $\left(2-\frac{t}{\Delta-a}\right)q+\frac{\Delta}{\Delta-a}r = 1$ has the $q$-intercept that is as large as $\frac{1}{2}$ and the $r$-intercept that is as large as $\frac{1}{\Delta}$. Also, $\frac{\Delta - a_v}{2\Delta - a_v - t_v} - \frac{1}{\Delta+1} = \frac{{\Delta}^2 - a_v\Delta - \Delta + t_v}{(2\Delta-a_v-t_v)(\Delta+1)} = \frac{{\Delta}(\Delta-a_v) - (\Delta - t_v)}{(2\Delta-a_v-t_v)(\Delta+1)} \ge \frac{\Delta - (\Delta-t_v)}{(2\Delta-a_v-t_v)(\Delta+1)} \ge 0$. Moreover, $d \ge 1$. Hence, if we let $ W_1 =  \{(q, r) \colon 2q + \Delta r > 1,  q > \frac{1}{\Delta+1}, \ r < \frac{\Delta - 1}{\Delta}q \}$, then $\Omega_1 \subseteq W_1$. \vskip 0.1in

Case 2) $S_{AB} = \emptyset$: \\
We let $a_v = \deg_{S_B}(v)$ for any $v \in S_B$. Then $a_v \le \Delta - 1$.
Now $r$ and $q$ must satisfy that (1) $a_v  > (1-q)\Delta $ and (2) $(1-q)a_v > (1-q-r)\Delta$. If we let $W_2 = \{(q, r) \colon q > \frac{1}{\Delta}, \  q+ \Delta r > 1\}$, then $\Omega_2 \subseteq W_2$.  \vskip 0.15in

Now we can see that $\{(q, r) \colon q, r > 0 \} - (W_1 \cup W_2)$ is the epidemic region for $T_{\Delta}$, and it is contained in $\Omega_G$. \qed
\end{pf}
\vskip 0.2in

The next result is that if an infinite regular graph contains an infinite regular tree, then the epidemic region for the graph is just the same as the epidemic region for an infinite regular tree. We consider a rooted tree $RT_{\Delta}$ defined as follows: the root $x_0$ has $\Delta-1$ children $x_1^1, x_1^2, \cdots, x_1^{\Delta}$, and each child has $\Delta$ neighbors, and so on. In other words, $T$ is an infinite rooted tree, where the root has degree ${\Delta-1}$ and all the rest of the vertices have degree $\Delta$.

\vskip 0.2in

\begin{thm} If an infinite $\Delta$-regular graph $G$ contains $RT_{\Delta}$ as a subgraph, then the epidemic region for $G$ is the same as the epidemic region for the tree $\Omega_{T_{\Delta}}$. 
\end{thm}
\vskip 0.2in
\begin{pf}
Let $H$ be a subgraph that is isomorphic to $RT_{\Delta}$ in $G$. Let $x_0$ be the root of $H$. For any vertex $x$ of $H$, the (induced) subtree starting at $v$ is in fact isomorphic to $H$. Therefore, for any finite subset of $V(G)$, say $C$, the remaining graph $G-C$ contains a vertex $v$ of $H$, and it contains a subtree isomorphic to $H$ with $v$ as the root. Also, note that an infinite component of $T_{\Delta} - C'$ for any finite set $C' \in  V(T_{\Delta})$ contains a subtree isomorphic to $RT_{\Delta}$ with a root $v'$ for some $v'$. Therefore, we choose each blocking structure for $(q, r)$ in $H - C$ exactly the same way as in $T_{\Delta} - C'$ for $(q, r)$. \smallskip

Now the epidemic region $\Omega_G^c$ contains $\Omega_{T_{\Delta}}^c$, and therefore $\Omega_G \subseteq \Omega_{T_{\Delta}}$. But by Theorem \ref{tree}, $\Omega_{T_{\Delta}} \subseteq \Omega_G$. Therefore, $\Omega_G  = \Omega_{T_{\Delta}}$.  \qed
\end{pf}

\vskip 0.25in
\begin{lem}\label{b}
Let $G$ be an infinite $\Delta$-regular graph with even $\Delta$. If for every finite subset $C$ of $V(G)$, 
\begin{enumerate} 
\item there are two disjoint non-empty subsets $S_B$ and $S_{AB}$ of $V(G) - C$ such that  
\begin{enumerate}
\item $\deg_{S_B}(v) \ge \frac{\Delta}{2}$ for every $v \in S_{AB}$
\item $\deg_{S_B}(u) + \deg_{S_{AB}}(u) = \Delta$ for every $u \in S_{B}$
\end{enumerate}
and
\item $G - C$ has a $\frac{\Delta}{2}$-regular subgraph,
\end{enumerate}
 then the epidemic region $\Omega_G$ is a subset of $\{(q, r) \colon q, r > 0 \} - (\Omega_i \cup \Omega')$ for some $1 \le i \le \Delta - 1$, where $$\Omega_i = 
 \left\{(q, r) \colon r < \frac{1}{2}q, \ \ 2q + \frac{\Delta}{\Delta - i}r > 1, \ \ q > \frac{\Delta-i}{2\Delta - i} \right\}$$
 $$\Omega' =  \left\{ (q, r) \colon q + 2r > 1, \ \ q > \frac{1}{2} \right\} .$$ 
 \end{lem}
 \vskip 0.15in
 
\begin{pf}
We show that for any finite subset $C$ of $V(G)$ and for any $(q, r)$ in $\Omega_1 \cup \Omega_2$, there is a blocking structure. \vskip 0.15in
We have two cases. \\

Case 1) Let $(q, r)$ be in $\Omega_i$, where $i = \min_{v \in S_B} \deg_{S_B}(v)$. We use $S_B$ and $S_{AB}$ guaranteed in condition 1 of the statement of Lemma \ref{b}. Now for any $u \in S_{AB}$, $\deg_{S_B}(u) \ge \frac{\Delta}{2} > \frac{r}{q}\Delta$ since $r < \frac{1}{2}q$. Now for any $v \in S_B$, let $a = \deg_{S_B}(v)$ and $b = \deg_{S_{AB}}(v)$, where $a + b = \Delta$. Since $S_B$ is non-empty, $a \ge 1$. Now we obtain the following for $1 \le a \le \Delta-1$: (1) $(1-q)a+qb - (1-q-r)\Delta = (\Delta - a)[2q+\frac{\Delta}{\Delta-a}r - 1] > 0$ and (2) $a+qb-(1-q)\Delta = q(\Delta+b)-(\Delta-a) > 0$.  Note that if $a' > a$, then $q$ and $r$ satisfy (1) and (2) after we replace $a$ by $a'$. We let $i = \min_{v \in S_B} \deg_{S_B}(v)$. Since $S_B \in G - C$, $1 \le i < \Delta$. \vskip 0.1in

Case 2) Let $(q, r)$ be in $\Omega'$. We let $S_{AB} = \emptyset$ and $S_B = V(H)$, where $H$ is a $\frac{\Delta}{2}$-regular subgraph guaranteed in condition 2 of the statement of Lemma \ref{b}. Since $S_{AB}$ is empty the first condition in the definition of a blocking structure is automatically satisfied. Now using the same notation as Case 1 we have $a = \frac{\Delta}{2}$ and $b = 0$ for every vertex $v \in S_B$. We check the second condition in the definition of a blocking structure: $(1-q)\frac{\Delta}{2} - (1-q-r)\Delta = \frac{\Delta}{2}(q+2r-1) > 0$. The third condition for being a blocking structure is satisfied because $a - (1-q)\Delta = \Delta(q-\frac{1}{2}) > 0$. 
 \vskip 0.1in
 
 By Theorem \ref{blocking} every $(q, r)$ in $\Omega_i \cup \Omega'$ is non-epidemic in $(G, q, r)$. \qed
 \end{pf}

\vskip 0.15in

\begin{cor}\label{c} If for every finite set $C$ there are two disjoint sets $S_B$ and $S_{AB}$ as in Lemma \ref{b} such that $\deg_{S_B}(v) = \deg_{S_{AB}}(v) = \frac{\Delta}{2}$ for every $v \in S_B$, then the epidemic region of $G$ is contained in $\Omega_{L_{\Delta}}$.
\end{cor}

\vskip 0.15in

\begin{pf}
The given condition implies that $i = \frac{\Delta}{2}$ in the proof of Lemma \ref{b}. Then,
$$\Omega_{\frac{\Delta}{2}} \cup \Omega' = \left\{(q, r) \colon r < \frac{1}{2}q, \ \ 2q + \frac{\Delta}{\Delta - \frac{\Delta}{2}}r > 1, \ \ q > \frac{\Delta-\frac{\Delta}{2}}{2\Delta - \frac{\Delta}{2}} \right\} \cup \left\{ (q, r) \colon q + 2r > 1, \ \ q > \frac{1}{2} \right\}$$
$$= \left\{(q, r) \colon r < \frac{1}{2}q, \ \ 2q+2r > 1, \ \ q > \frac{1}{3} \right\} \cup \left\{ (q, r) \colon 1+2r > 1, \ \ q > \frac{1}{2} \right\}.$$
Therefore, the epidemic region of $G$ is
$$ \left\{(q, r) \colon q, r > 0 \right\} - \left(\Omega_{\frac{\Delta}{2}} \cup \Omega' \right) =   \left\{(q, r) \colon q \le \frac{1}{2}, \ \ r \ge \frac{1}{2}q \right\} \cup \left\{ (q, r) \colon r \le \frac{1}{2}q, \ \ 2q + 2r \le 1 \right\}.$$ 
Note that the resulting set is the same as $\Omega_{L_{\Delta}}$ \cite{Immorlica07}.
\qed
\end{pf}

\vskip 0.15in

We denote $HL_{\Delta}$ a thick half line graph for even $\Delta$. In other words, the vertex set of $HL_{\Delta}$ can be described as $\mathbf{N} \times \{1, 2, \cdots, \Delta/2\}$. There is an edge between $(k, i)$ and $(l, j)$ if and only if $|k - l| = 1$. In $HL_{\Delta}$ every vertex $(k, i), \ k \ge 2$ has degree $\Delta$, and every vertex $(1, i)$ has degree $\Delta/2$. \vskip 0.15in 

We will construct a family of infinite $\Delta$-regular graphs. The family contains the thick line graph as a special case. \smallskip
 For $N \ge 2$ and an even $\Delta$ we consider $N$ copies of $HL_{\Delta}$: $HL(1), HL(2), \cdots, HL(N)$. Let $V(i)$ be the vertices of $(1, 1), (1, 2), \cdots, (1, \Delta/2)$ of each copy $HL(i)$. We identify  $t_{ij}$ vertices in $V(i)$ with $t_{ji}$ vertices in $V(j)$ for $i \neq j$ for some $t_i \ge 0$ and for some $t_{ij} \le \frac{\Delta}{2}$, where $t_{ij} = t_{ji}$.  We add finitely many vertices to the above graph and add appropriately many edges such that every vertex has degree $\Delta$. Note that for $N = 2$ if $v_{12} = v_{21} = 0$ and we add a perfect matching between $V(1)$ and $V(2)$, then the resulting graph is isomorphic to $L_{\Delta}$.
 
\begin{thm}
The epidemic region for any graph from the construction is the same as $\Omega_{L_{\Delta}}$. 
\end{thm} \vskip 0.1in

\begin{pf}
We denote $G$ an infinite $\Delta$-regular graph from the construction and $\Omega_G$ the epidemic region for $G$. We let $C$ be a set of vertices of $G$ consisting of $V(i), \ i = 1, 2, \cdots, N$ and the added vertices from the construction. Hence, $C$ is finite. Starting from the initial state where every vertex takes the strategy $B$, the vertices in $C$ take new strategy $A$. Now we consider $(G, q, r)$ game with every vertex taking its best strategy. \vskip 0.1in
We name group $g_m^i$ each group of vertices $(i, 1), (i, 2), \cdots , (i, \Delta/2)$ in $HL(m)$. We determine an order of moves of the vertices with strategy $B$ as follows: $g_1^1, g_1^2, \cdots, g_1^N, g_2^1, \cdots , g_2^N, g_3^1, \cdots$. Note that in each group any order of the $\Delta/2$ vertices works. Then at each stage the payoffs of strategies $A, B, AB$ are $\frac{\Delta}{2}(1-q), \frac{q\Delta}{2}$, and $\frac{\Delta}{2}-r\Delta$, respectively. (Note that it has the same pattern as in the thick line graphs.) Hence, for $(q, r)$ with $q \le \frac{1}{2}$ and $r \ge \frac{1}{2}q$ the best strategy of a vertex at its turn is $A$.  \vskip 0.1in

On the other hand, for $(q, r)$ with $r \le \frac{1}{2}q$ and $2q + 2r \le 1$ the best strategy of a vertex at its turn (in the same order as above) is $AB$. As it is described in \cite{Immorlica07}, the payoff of strategies $A$, $B$, and $AB$ are $\frac{1-q}{2}\Delta$, $q\Delta$, and $\left( \frac{q+1-q}{2}\Delta \right) - r\Delta$, respectively. Hence, if $AB$ has the best payoff, then $\frac{\Delta}{2} - r\Delta \ge \frac{1-q}{2}\Delta$ and $\frac{\Delta}{2} - r\Delta \ge q\Delta$. The inequalities are simplified to $r \le q/2$ and $2q+2r \le 1$. We conclude that $\Omega_G$ contains $\{(q, r) \colon r \le \frac{1}{2}q, 2q + 2r \le 1 \} \cup \{ (q, r) \colon q \le \frac{1}{2}, r \ge \frac{1}{2}q \}$. \vskip 0.15in

To determine the complement of $\Omega_G$ we consider blocking structures. For any deletion of a finite set of vertices $C$ the remaining graph still contains a thick half line. Then we find the same blocking structures used for determining the complement of $\Omega_{L_{\Delta}}$. \vskip 0.15in

In $L_{\Delta}$, two blocking structures are used: A $\frac{\Delta}{2}$-regular subgraph $K_{\frac{\Delta}{2}, \frac{\Delta}{2}}$ satisfies condition 2 in the statement of Lemma \ref{b}. Also, $G - C$ contains a thick half line as an induced subgraph by a set of vertices $S = \{(k, i) \colon k \ge m, 1 \le i \le \frac{\Delta}{2} \}$ for some $m \ge 1$ from a copy of $HL_{\Delta}$. We partition $S$ into $S_{B}$ and $S_{AB}$. $S_{B} = \{ (k, i) \colon k = m, m+2, m+4, m+6, \cdots \}$ and $S_{AB} = \{ (k, i) \colon k = m+1, m+3, m+5, \cdots \}$. Now $S_{B}$ and $S_{AB}$ satisfy condition 1 in the statement of Lemma \ref{b}. Moreover, since $\deg_{S_B}(v) = \deg_{S_{AB}}(v) = \frac{\Delta}{2}$ for every $v \in S_B$, by Corollary \ref{c} $\Omega_G$ is contained in $\Omega_{L_{\Delta}}$. \qed
\end{pf}

\begin{figure}[tb!]
\centerline{\includegraphics[width=7cm]{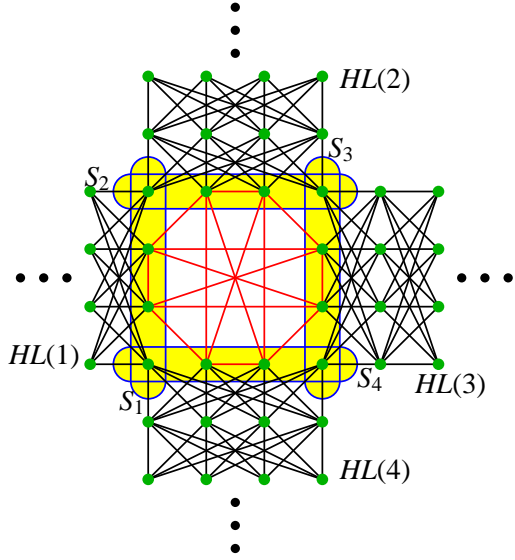}}
\caption{ $G_{8,2}$ (in Corollary \ref{cor})
}
\label{FIG_G_8_2}
\end{figure}

\begin{cor}\label{cor}
For every even $\Delta > 2$, there are infinitely many infinite $\Delta$-regular graphs whose epidemic region is the same as $\Omega_{L_{\Delta}}$. 
\end{cor}\vskip 0.1in

\begin{pf}
We let $k \ge 2$. We construct $G_{\Delta, k}$ for even $\Delta$ as follows. \\
Let $S$ be a set of $\Delta/2$ elements. We label the elements of $S$ as $x_1, x_2, \cdots, x_{\frac{\Delta}{2}}$. We consider $2k$ copies of $S$. Let $S_m$ be the $m$-th copy of $S$. Let $x_1^{m}$ and $x_{\frac{\Delta}{2}}^{m}$ be the first and the last elements respectively in $S_m$. We identify $x_1^{m}$ and $x_{\frac{\Delta}{2}}^{m-1}$. In other words, $S_{m-1} \cap S_{m} = \{ x_1^{m} \}$. We make the infinite half-line $HL(m)$ with degree $\Delta$ starting at $S_m$, where the elements of $S_m$ have degree $\frac{\Delta}{2}$. 
Now we consider the $2k\left(\frac{\Delta}{2}-2\right)$ vertices in $\bigcup_{m = 1}^{2k}S_{m} - \bigcup_{m = 1}^{2k} \left\{x_1^{m}, x_{\frac{\Delta}{2}}^{m} \right\}$. \\
It is well-known that there is a connected $\frac{\Delta}{2}$-regular graph $G'$ on $2k\left(\frac{\Delta}{2} - 2\right)$ vertices. We consider the following graph $G_{\Delta, k}$ obtained by adding the edges of $G'$ to the union of $S_m$s for $m = 1, 2, , ..., 2k$. In other words, $G_{\Delta, k}= \bigcup_{m = 1}^{2k}{S_m} \cup E(G')$. Now the epidemic region of $G_{\Delta, k}$ is the same as $\Omega_{L_{\Delta}}$. \qed
\end{pf} \vskip 0.25in

\section{Discussion}
The simplest non-trivial infinite regular graph is when $\Delta$ is 2. There is only one possible such graph: infinite line $L_2$, which attains the smallest and the largest possible epidemic region simultaneously. For even $\Delta > 2$ there are many non-isomorphic infinite $\Delta$-regular graphs including $T_{\Delta}$ and $L_{\Delta}$. We showed that the regular tree has the smallest epidemic region, and we showed that any infinite regular graph containing the infinite regular tree must have the minimum epidemic region. \\

Our prediction is that for any infinite $\Delta$-regular graph $G$, $\Omega_{T_{\Delta}} \subseteq \Omega_G \subseteq \Omega_{L_{\Delta}}$. 
In fact for bigger epidemic regions, the maximum possible epidemic region known so far is $\Omega_{L_{\Delta}}$. However, we still do not know whether $\Omega_{L_{\Delta}}$ has the maximum epidemic region among all infinite $\Delta$-regular graphs. We conjecture that $\Omega_{L_{\Delta}}$ is the maximum region among all infinite $\Delta$-regular graphs. \\

Along the lines of the results, here are various directions for further research. 
One way is that we can apply the setting in a contagion game for non-regular graphs, where the (total) payoffs for a vertex $v$ are $(\alpha+\beta)(1-q), (\deg(v)-\alpha)q$, and $(\alpha+\beta)(1-2q)+q \deg(v)-c$ for strategies $A, B$, and $AB$, respectively. 
Here each of $\alpha$, $\deg(v)-\alpha-\beta$, and $\beta$ is the number of neighbors with $A, B$, and $AB$, respectively. 
It is reasonable to assume that the cost $c$ is constant so that a vertex with a high degree tends to adopt a bilingual strategy because $r_v = {c}/{\deg(v)}$. 
For non-regular graphs, an epidemic region can be defined in a similar way. But in this case it is in the $(q, c)$-plane instead of the $(q, r)$-plane.  \medskip

Another possible model reflecting reality more is that we consider a finite graph with $N$ vertices, where $N$ is the population of a network. 
It is not clear whether an epidemic behavior on infinite graphs has the same characterization as the limit of the behavior on finite graphs with $N$ vertices as $N \to \infty$. 
A corresponding definition is that we say $A$ is epidemic on a graph with $N$ vertices if there is a finite set of $S$ with size $f(N)$ and a finite sequence of vertices in $V(G) - S$ for playing the game, $v_1, v_2, v_3, \cdots, v_M$, satisfying the following conditions:
\begin{enumerate}
\item $\lim_{N \to \infty} \frac{f(N)}{N} = 0 $, and 
\item for every $v$ in $V(G) - S$ there is $k$ such that (1) $v = v_k$ and (2) at the $k$-th turn in the sequence its best strategy for the vertex is $A$ when having started with $A$ for the vertices of $S$ while other vertices have $B$.
\end{enumerate} \vskip 0.15in

Is there an infinite family of graphs $\{ G_n \colon |V(G_n)| = n, n = 1, 2, 3, \cdots \}$ such that for each corresponding finite set $S_n$ making $G_n$ epidemic $f(n)$ is constant? In other words, $f(n) = o(1)$? Also, we can ask how large $f(N)$ could be for an $N$-vertex graph to be epidemic. How is $f(N)$ related to structure of graphs?

\vskip 0.15in

\section*{Acknowledgments}
This work was supported by GIST Research Institute (GRI) grant funded by the GIST in 2018. \vskip 0.05in

\section*{References}
\bibliography{bilingual}

\end{document}